\documentclass{iau} 
\usepackage{graphicx,natbib}

\usepackage[version=3]{mhchem}

\def\apj{ApJ}%
\def\apjl{ApJ}%
\def\apjs{ApJS }%
\def\aap{A\&A}%
\def\mnras{MNRAS}%
\def\jcp{J. Chem. Phys.}%

\title[Surface astrochemistry
] 
{Surface astrochemistry: a computational chemistry perspective
}

\author[H.M. Cuppen et al.]   
{H.M.~Cuppen$^1$, A.~Fredon$^1$,  T.~Lamberts$^2$,
 E.M.~Penteado$^1$, M.~Simons$^{1}$ \and C.~Walsh$^{3,4}$ }

\affiliation{$^1$Radboud University, Institute for Molecules and Materials \\ email: {\tt h.cuppen@science.ru.nl} \\[\affilskip]
$^2$Computational Chemistry Group, Institute for Theoretical Chemistry, University of Stuttgart, Pfaffenwaldring 55, 70569 Stuttgart, 
 Germany\\[\affilskip]
$^3$School of Physics and Astronomy, University of Leeds, Leeds LS2 9JT, UK\\[\affilskip]
$^4$Leiden Observatory, Leiden University PO Box 9513, 2300 RA Leiden, The Netherlands\\[\affilskip]}

\pubyear{2017}
\volume{332}  
\jname{Astrochemistry VII -- Through the Cosmos from Galaxies to Planets}
\editors{Maria Cunningham, Tom Millar and Yuri Aikawa, eds.}
\begin{document}

\maketitle

\begin{abstract}
Molecules in space are synthesized via a large variety of gas-phase reactions, and reactions on dust-grain surfaces, where the surface acts as a catalyst. Especially, saturated, hydrogen-rich molecules are formed through surface chemistry. Astrochemical models have developed over the decades to
understand the molecular processes in the interstellar medium, taking into account grain surface chemistry. However, essential input information for gas-grain models, such as binding energies of molecules to the surface, have been derived experimentally only for a handful of species, leaving hundreds of species with highly uncertain estimates. Moreover, some fundamental processes are not well enough constrained to implement these into the models.

The proceedings gives three examples how computational chemistry techniques can help answer fundamental questions regarding grain surface chemistry. 
\end{abstract}

\firstsection 
\section{Introduction}
Close to 200 molecules have been identified in interstellar space (see {\em http://www.astro.uni-koeln.de/cdms/molecules}). Among these, approximately 50 are classified as complex organic molecules (COMs), i.e., comprised of six atoms or more. COMs are generally observed in the gas phase, although it is widely accepted that saturated COMs are formed primarily through reactions on the surface of icy dust grains. This is where gas-phase species accrete, meet and react to form saturated molecules. The new generation of telescopes provides us with more information about that spatial extent, fractionation, and correlation with other COMs.  The analysis of these data will require much more elaborate, and more diverse, gas-grain astrochemical models than have been developed so far \citep{Cuppen:2017}. However, essential input information for gas-grain models, such as binding energies of molecules to the surface, have been derived experimentally only for a handful of species, leaving hundreds of species with highly uncertain estimates. Moreover, some fundamental processes are not well enough constrained to implement into the models.

For instance, the precise mechanism to form COMs with a \ce{C-C} bond is unknown and several different mechanisms have been suggested over the past decades. \citet{Charnley:2009} applied a simple atom addition mechanism where species such as methanol can react with atomic carbon. Another mechanism was proposed by \citet{Garrod:2006a} and is more efficient at elevated temperatures. Cosmic-ray-induced photons can dissociate \ce{H2CO} and \ce{CH3OH}, creating functional-group radicals such as \ce{CH3} and \ce{CH3O}, on/within the ice mantle. At low temperatures, these radicals are hydrogenated again, but as the temperature increases to above $20-30$~K, the residence time of H atoms on the surface decreases substantially while at the same time radicals become mobile. Radical-radical association reactions become competitive with hydrogenation of radicals. 

The detection of acetaldehyde \ce{CH3CHO}, dimethyl ether \ce{CH3OCH3}, methyl formate \ce{CH3OCHO}, and ketene \ce{CH2CO} in the prestellar core L1689B  suggest that COM synthesis has already started at the prestellar stage and suggests that there should be a viable grain surface reaction route that is possible at low temperatures and without the need for extensive irradiation \citep{Bacmann:2012}.
\citet{Woods:2013} proposed a mechanism for the formation of COMs through recombination of two formyl radicals (HCO). Recently, \cite{Fedoseev:2015}  confirmed this picture by showing experimentally that the reaction between two HCO radicals yields glyoxal (HC(O)CHO) at 13 K; sequential hydrogenation with two or four hydrogen atoms produces glycoaldehyde and ethylene glycol, respectively.
HCO radicals are formed in this case by CO hydrogenation and not by UV photolysis. Moreover, it showed that radical-radical recombination reactions are not only efficient at temperatures higher than 20--30 K, when radicals are commonly assumed to become mobile \citep[e.g.,][]{Garrod:2013, Butscher:2015}. 

In this proceedings, we will give three examples how computational chemistry techniques can help answer fundamental questions regarding grain surface chemistry. The low $T$ COM formation route will be explored in more detail using a microscopic kinetic Monte Carlo (KMC) model. By applying a sensitivity analysis on binding energy data, more information on  the relative importance of the different COM formation routes can be obtained.  Finally, the role of exothermicity upon reaction for subsequent diffusion and desorption will be quantified using Molecular Dynamics simulations.

\section{COM formation through HCO recombination without diffusion}
The hypothesis of the low temperature COM formation route is that two formyl radicals (\ce{HCO}) form in close proximity and recombine to from glyoxal (\ce{HC(O)CHO}) \citep{Fedoseev:2015,Chuang:2016}. Through continued hydrogenation, glyoxal will consecutively be turned into glycoaldehyde (\ce{HC(O)CH2OH}) and ethylene glycol (\ce{H2C(OH)CH2OH}). Finally, recombination of a formyl and a methoxy radical leads to the formation of methyl formate (\ce{CH3OCHO}). Formyl is formed through the well-known hydrogenation sequence of CO to methanol \citep{Watanabe:2002,Fuchs:2009} and hence without the need of UV photolysis or cosmic rays as external triggers. During CO freeze-out the gas phase H-atom abundance is similar to the CO gas phase abundance and there will be an insufficient number of H atoms to fully hydrogenate CO to \ce{H2CO} and \ce{CH3OH}, leaving a relatively high HCO surface abundance. Statistically these can be formed sufficiently close to recombine without the need of long scale diffusion. 
 
\begin{figure*}[t!]
     \centering
     \includegraphics[width=\textwidth]{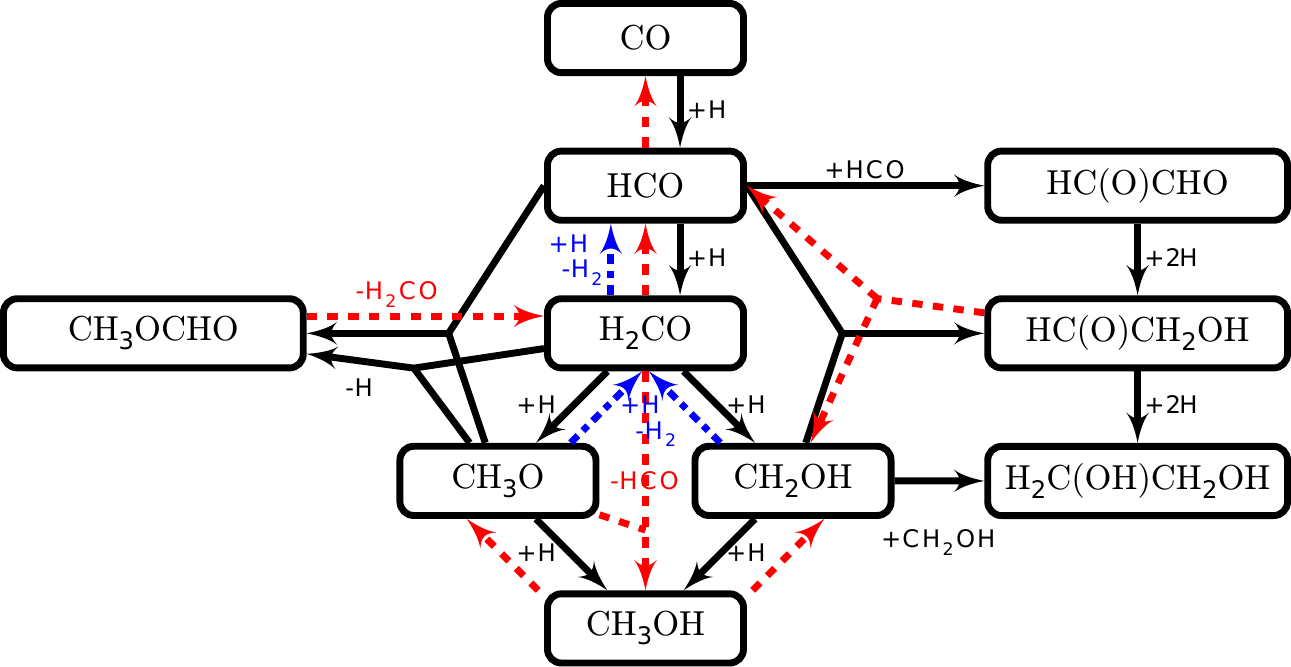}
     \caption{The new reaction network proposed in this paper. Solid lines indicate addition reactions whereas red dashed lines mark photo-dissociation reactions and blue dash-dotted lines are thermal dissociation reaction.}
     \label{fig:scheme}
\end{figure*}

This hypothesis has been shown to work in the laboratory \citep{Fedoseev:2015,Chuang:2016}, but here fluxes are necessarily high to enable molecule formation within the timescale of a few hours. To put the hypothesis really to the test, we simulate COM formation under interstellar conditions using  a microscopic KMC model. We have previously been successful with this model to simulate both laboratory and ISM conditions for the formation of methanol \citep{Fuchs:2009,Cuppen:2009}. The main strength of microscopic simulation methods is that the location of individual species is traced throughout the simulation. Our study of hydrogenation of CO \citep{Cuppen:2009} showed that a gradient mantle composition builds up during catastrophic CO freeze-out. This is in agreement with ice observations of CO \citep{Penteado:2015}. 

In the past years, several codes have been developed that introduce multiple phases representing different layers within the ice \citep{Taquet:2012,Vasyunin:2013,Furuya:2016}. However, the reaction rates are still calculated assuming a diffusive mechanism, by either rate equations or a macroscopic KMC routine. Crucial to the low temperature formation mechanism is that two HCO radicals form in close vicinity of each other and can react without the need for diffusion.  A microscopic Kinetic Monte Carlo model has access to positional information of all species and does not make assumptions on the diffusive character of a reaction.  It can therefore be used to estimated the relative importance of the different COM formation channels. 

\begin{figure*}[t!]
        \centering
        \includegraphics[width=0.9\textwidth]{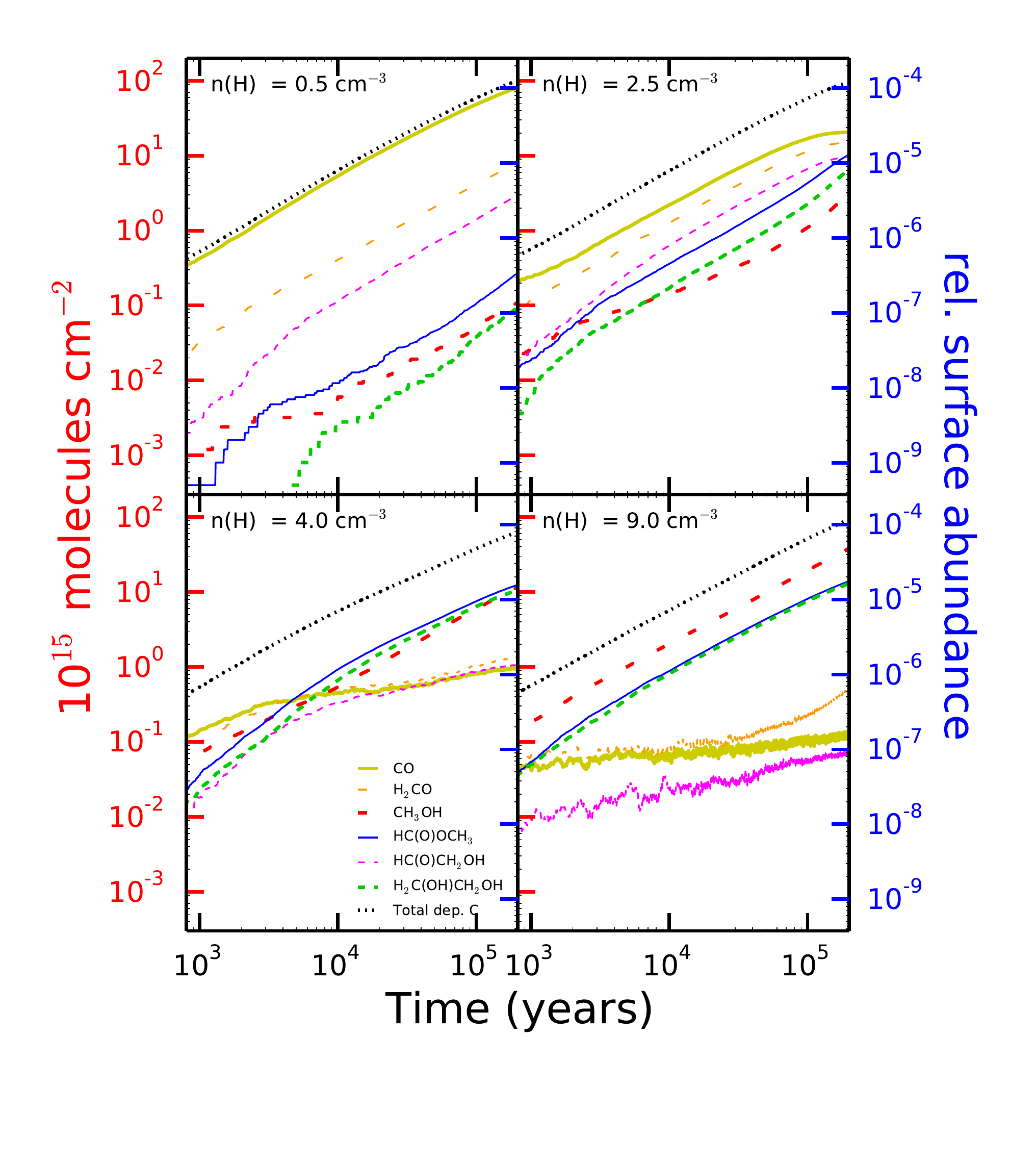}
        \caption{Time evolution of the grain surface composition for four different levels of initial hydrogen abundances: 0.5, 2.5, 4.0, and 9.0 cm$^{-3}$. The abundance of gas-phase carbon monoxide is 10.0 cm$^{-3}$. The temperature of the grain is 10 K. The relative surface abundance is given with respect to $n_\text{H}$. }
        \label{fig:timeevolution}
\end{figure*}

Figure \ref{fig:timeevolution} shows four simulation results for dark molecular clouds conditions with different hydrogen abundances. The simulated abundances are $n(\ce{H}) = 0.5, 2.5, 4.0,$ and 9.0~cm$^{-3}$, respectively. The grain surface temperature is 10~K and the initial CO abundance is $n_\text{initial}(\ce{CO}) =10$~cm$^{-3}$ in all cases. The gas phase abundance of atomic hydrogen is assumed to be in steady state in the gas phase and to remain constant throughout the simulation. The initial CO gas phase abundance is assumed to hold the total CO reservoir and is diminished throughout the simulation accounting for CO on the surface, and its products. The left-hand, vertical axes (in red) give the build-up in number particles per surface area whereas the right-hand axes (in blue) denote the grain surface abundance with respect to $n_\text{H}$. The total simulated time is $2\times 10^5$~years.  The upper left panel is a the result of simulations with a low H:CO ratio. The grain mantle is mainly composed of CO with \ce{H2CO} as second most abundant species. In these conditions the formation of HCO is relatively rare and two HCO radicals are hardly formed in close proximity. Indeed, the low amount of glycoaldehyde formed indicates that little HCO recombinations have occurred. However, enough hydrogen atoms are present to further hydrogenate HCO to \ce{H2CO}. A slow decrease in the mantle build-up (gray line) can be observed at approximately $10^5$ years, due to CO depletion from the gas phase. 

Results of simulations with a slightly higher gas-phase hydrogen abundance are shown in the upper right panel. The grain mantle now features less CO and \ce{H2CO} compared to the lower H:CO ratio. Because of the higher amount of hydrogen present in the gas phase, more HCO radicals are formed which can form \ce{HC(O)CHO} instead of hydrogenating further to yield \ce{H2CO}.  This is particularly true towards the end of the simulation, when the H:CO ratio increases further. Please notice that the overall grain mantle build-up in number of molecules is slightly lower than in the top left panel as indicated in the gray line. This is partly because it takes two molecules of CO to form \ce{HC(O)CHO} and its hydrogenation products whereas it only takes one CO molecule to form \ce{H2CO}. The red dotted line accounts for this difference as it represents the number of equivalent CO molecules per unit array. The difference in mantle build-up can be fully explained by this effect.

Upon increasing the H:CO ratio further, HCO radical combination starts to compete with further hydrogenation. For the H:CO = 0.4 results in the bottom left bottom panel, the lifetime of the formation rates of \ce{CH3OH} by further hydrogenation, \ce{H2C(OH)CH2OH} through HCO radical formation, and \ce{CH3OCHO} through reaction of \ce{CH3O} are of comparable magnitude. This can be concluded from the abundance of these three species. The red dotted curve in the graph further shows that the mantle build-up is hampered.  We will come back to this point later. For the highest H:CO ratio, hydrogenation wins from HCO radical recombination leading a \ce{CH3OH} dominated mantle with little \ce{H2C(OH)CH2OH} and \ce{CH3OCHO}.

\begin{figure}[t!]
    \centering
    \includegraphics[width=0.2\textwidth]{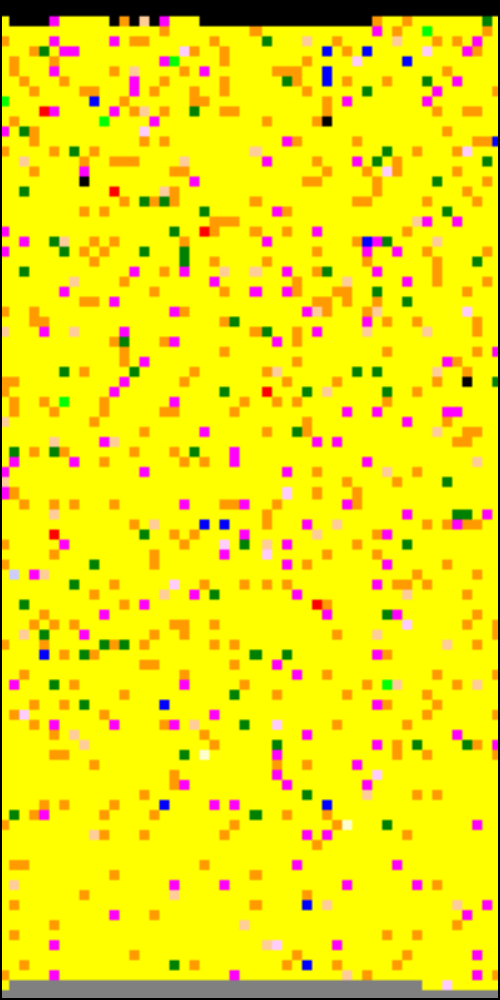}
    \includegraphics[width=0.2\textwidth]{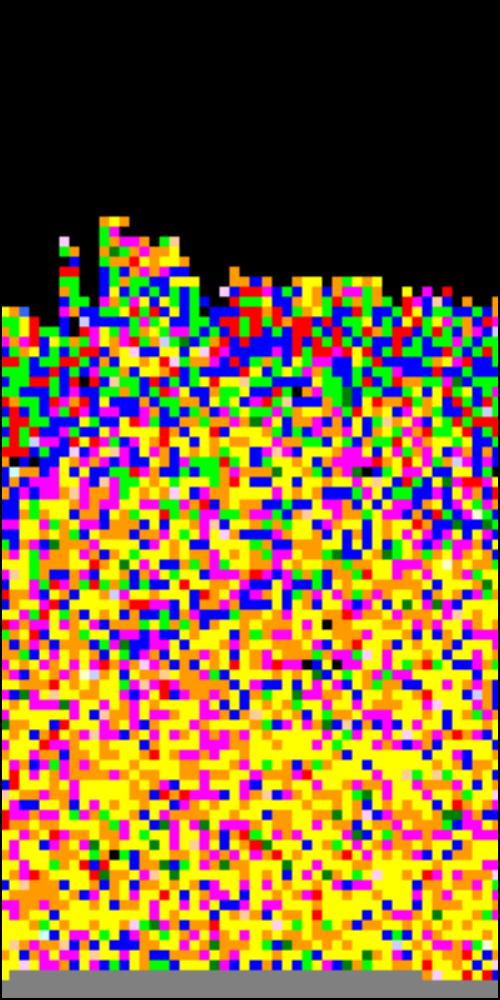}
    \includegraphics[width=0.2\textwidth]{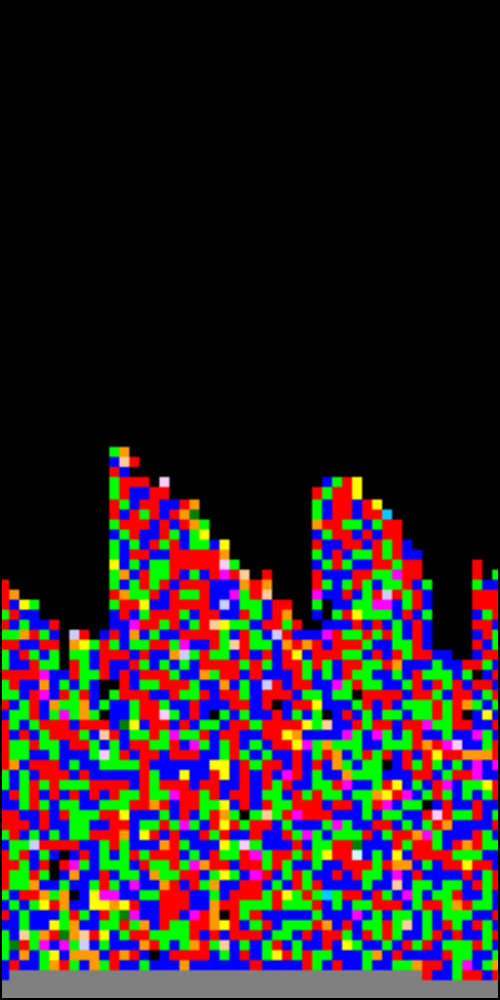}
    \includegraphics[width=0.2\textwidth]{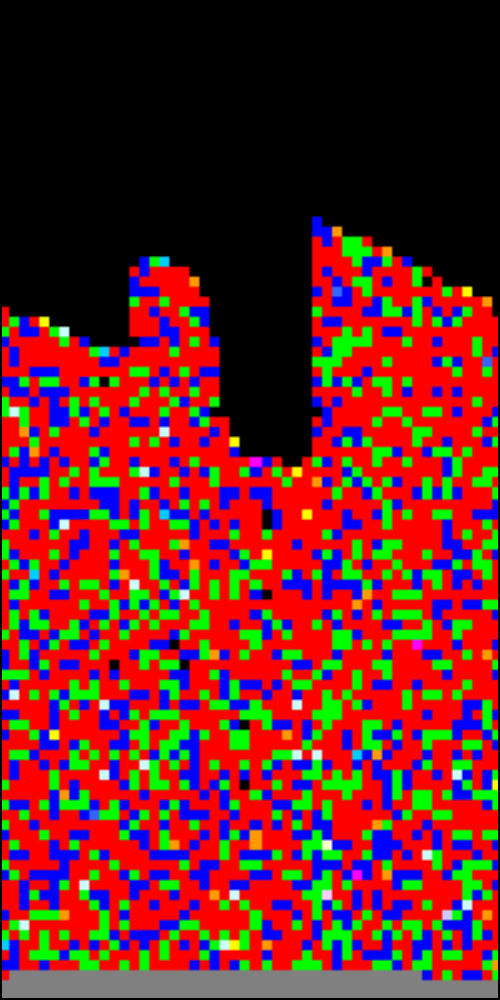}
    \caption{Cross-sections of the grain surface at $n(\text{H})=$ 0.5, 2.5, 4.0, and 9.0 cm$^{-3}$. Colour indications are equal to Figure \ref{fig:timeevolution}. Intermediates have slightly brighter colors than their immediate product. The grain is colored gray and the black parts are vacuum.}
    \label{fig:cross}
\end{figure}

Figure \ref{fig:cross} shows cross sections of the grain mantles at $t = 2\times 10^5$ years for the four simulations in Fig.~\ref{fig:timeevolution}. All panels show the grain in gray at the bottom of the image and the gas phase in black at the top. The different species that occupy the simulation lattice are represented by different colors. The same color coding is used as in Fig.\ref{fig:timeevolution}. The composition of the grain mantle reflects the earlier discussion on  Fig.~\ref{fig:timeevolution}. What is immediately evident is the influence of the H:CO ratio on the surface morphology and the mantle height. Some of the reduction in height is because of the formation of the C--C bond which leads to a reduction in the number of molecules as mentioned earlier. The reduction for the $n(\text{H})=$ 4.0 cm$^{-3}$ simulation result is real. This appears be accompanied by a more rough surface morphology. We believe this because for more rough surfaces CO and H are more likely to land on top of a protrusion, which has less favorable binding since species are only interacting with a few neighboring species. CO and H are hence more likely to desorb from these protrusions before they can react. Moreover, these protrusions also hinder the diffusion to stronger binding sites.

The KMC simulations show that complex organic molecules with a C--C bond can be formed at temperatures as low as 10~K and without the need of external radiation. The H:CO gas phase abundance ratio is crucial in determining the amount of COMs and their relative composition. According to \citet{Goldsmith:2005}, the atomic hydrogen abundance should be in the range $n(\ce{H})\approx2$--$6$ cm$^{-3}$ which is in our intermediate H:CO range with optimal COM production circumstances. The amount of COMs formed is more than expected and for future simulations we need to consider more destruction channels.

Finally, all results discussed previously take into account the exothermicity of the surface reactions. The excess energy that is released during the reaction can be used for diffusion and desorption. In our scheme, a reaction product can move around a few hops before it is thermalized. This allows the reactants to find a site with a stronger binding energy and it makes the ice formed  more compact. Simulations without exothermicity taken into account result in very fractal-like surfaces and show a substantial hampering in the ice build up. New species are less likely to accrete on these fractal-like surfaces and the binding energy in these structure is lower, enabling desorption of the reactants. This reinforces the conclusion that exothermicity is an important factor in the build up of ices \citep{Lamberts:2014I} .

\section{Sensitivity analysis: the relative importance of the different COM formation routes}
Next, we present a sensitivity analysis study for the grain surface part of gas-grain codes by looking
at the effect of uncertainties in the binding energy ($E\rm_{bind}$) of species to the grain. We use a similar approach to  \citet{Wakelam:2006} and \citet{Wakelam:2010B}.  The binding energy of a species determines the upper value of the temperature range at which species are
available for reactions on the grain surface. It also, indirectly, determines the onset temperature for surface
reactions, since in many models the diffusion barrier is considered to be a fixed fraction of the binding energy \citep{Hasegawa:1992, Ruffle:2000, Garrod:2006a, Cuppen:2009}.
The initial goal of this work was to obtain constraints on the binding energies of different surface species by comparing the simulation results against ice observations. As will become clear, it is hard to constraint binding energies in this way, but crucial insight into the reaction network is obtained.

We start by compiling an updated list of binding energies by reviewing current experimental data. This resulted in 
recommended values for the binding energies and their uncertainties. These uncertainties reflect experimental uncertainties as well as intrinsic variations due to a range of different binding sites present on the surface. For radicals species, limited data is available and here we used uncertainties of half the binding energy with a maximum of 500~K. Next, 10,000 simulations of a standard homogeneous dark cloud with constant
physical conditions are performed, where each simulation uses a set of binding energies that
is randomly picked from a Gaussian distribution considering the recommended binding
energies and their uncertainties. The rate equation model used is a two-phase chemistry model, since it treats the gas 
and solid phase; however, without using location information of ice species within the ice mantle. The chemical network used in the present work \citep[][and references therein]{Drozdovskaya:2014, Drozdovskaya:2015, Garrod:2008b, Walsh:2015} 
contains 190 surface species. More details can be found in \citet{Penteado:2017}.

\begin{figure}
  \begin{center}
     \centering
     \includegraphics[width=\textwidth]{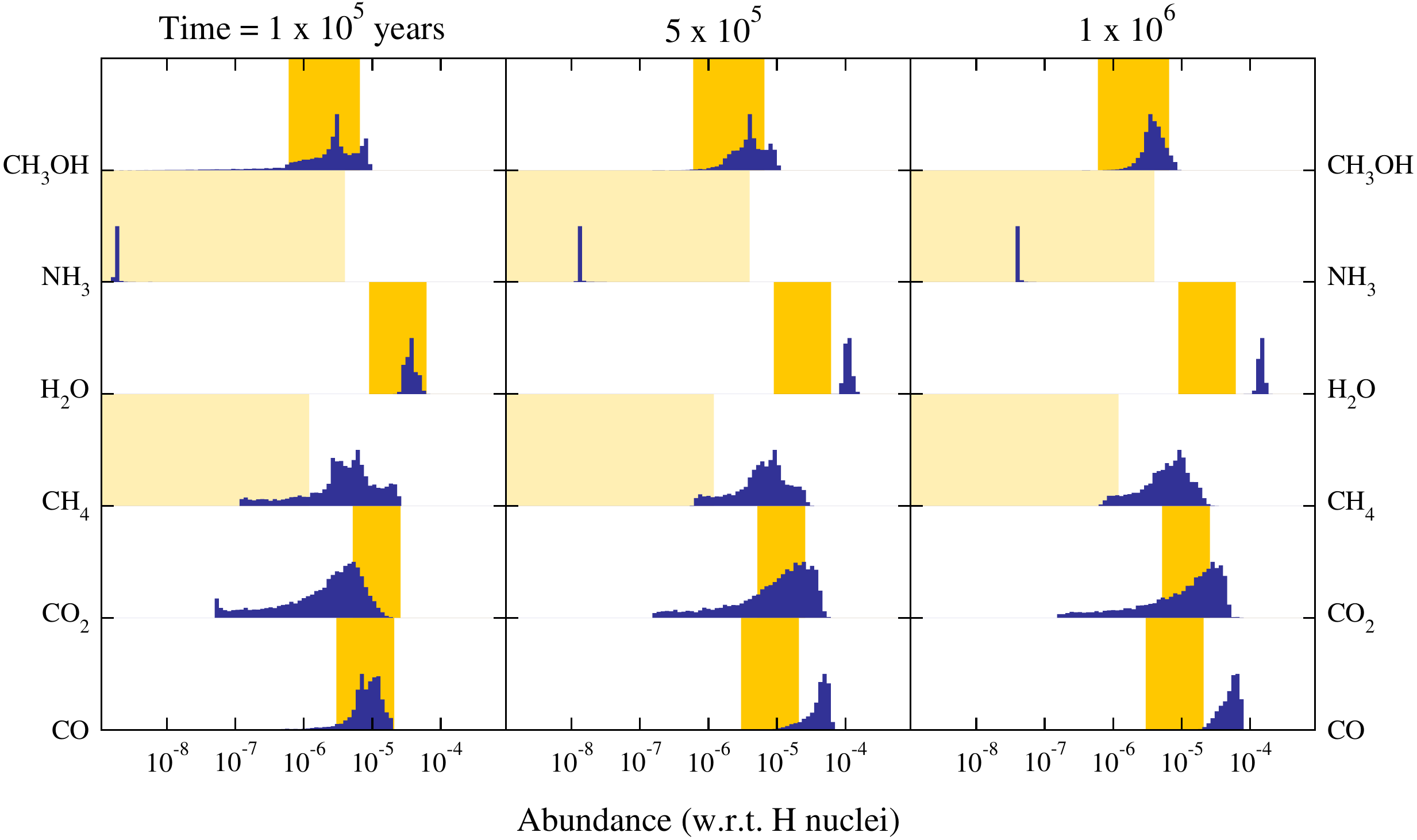}
      \caption{Distributions of simulated ice abundances (blue histograms)  for three different times (10$^5$ in the left, 
      5$\times$10$^5$ in the middle and 10$^6$ years in the right panels) compared to abundances derived from observations (yellow areas, light yellow for upper limits) of quiescent clouds and cores (see \cite{Boogert:2015} for references). Figure reproduced from \citet{Penteado:2017}.}
        \label{comp_obs}
  \end{center}
\end{figure}

The simulations result in a very broad distributions of the obtained ice abundances. Only \ce{H2O} has a narrow distribution, other species can vary over orders of magnitude. Figure~\ref{comp_obs} shows the distributions for \ce{H2O}, \ce{CO}, \ce{CH3OH}, \ce{NH3}, and \ce{CH4} at three different times. Although the full distribution can cover several orders of magnitudes, the dispersion in terms of full-width-half-maximum is significantly narrower ($\lesssim 1$ order of magnitude) for most species. We compare the obtained model abundances with observational abundances taken from a recent review by \cite{Boogert:2015} where
we have taken the background star observations as representative of observations of quiescent clouds and cores.  What is clear is the very large spread in observational abundances as indicated by the yellow areas in Figure~\ref{comp_obs}, from upper limits to relatively large values compared
with water ice observations. Inspection of Figure~\ref{comp_obs} shows that the best agreement is obtained 
at relatively early times of $10^5$ years. This coincides with dark cloud model results focusing on the gas phase, where $10^5$ years is considered as the early time of best overall agreement between 
models and observations for a large number of species.
The observations of CH$_4$ and NH$_3$ show only upper limits. Our model results for NH$_3$ fall well below this limit, where as a large fraction of models heavily overproduce CH$_4$. We come back to this later.

   \begin{figure}
   \begin{center}
     \centering
     \includegraphics[width=0.8\textwidth]{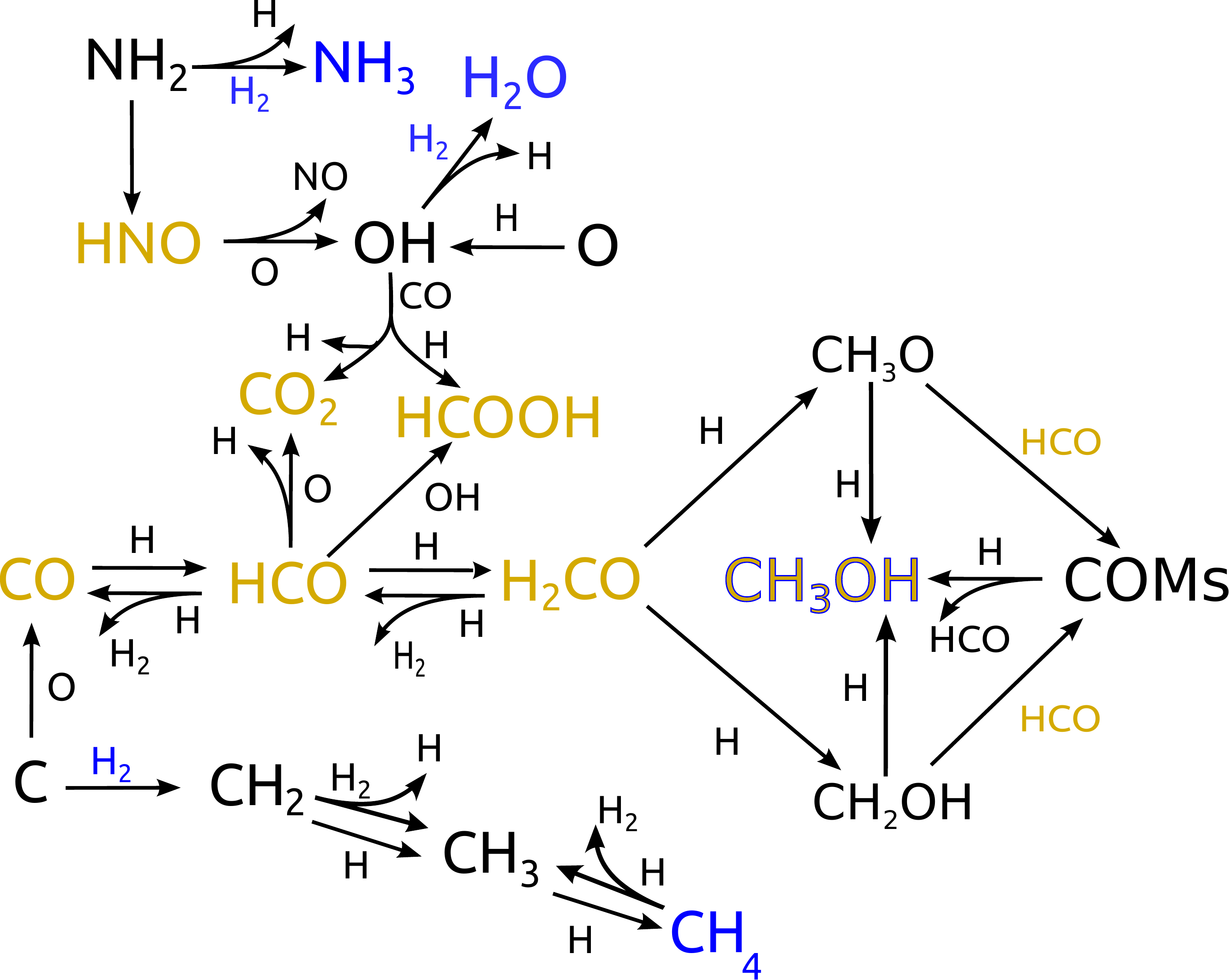}
      \caption{A limited surface network explaining the main results.  Figure reproduced from \citet{Penteado:2017}.}
        \label{network}
   \end{center}
   \end{figure}

A detailed analysis of the dependence on the different binding energies resulted in a limited surface network that can explain the main features of the results. This is presented in Figure \ref{network}. Crucial here is whether CO hydrogenation or \ce{CH4} formation wins, which is critically determined by the amount of \ce{H2} on the surface. For high \ce{H2} abundance, the reactions and species in blue dominate, otherwise the yellow reactions and species dominate. Methanol is both yellow and blue and its abundance does not correlate with CO, \ce{H2CO}, and other CO hydrogenation products. The reason for this, is that for high HCO abundance, intermediate species \ce{CH3O} and \ce{CH2OH} rather react with HCO than hydrogenate to form \ce{CH3OH}. \ce{CH3OH} is, in this case, mainly formed through destruction of more complex species. For lower HCO abundances, \ce{CH3OH} is formed directly through hydrogenation of \ce{CH3O} and \ce{CH2OH}. Close inspection of the network shows that the HCO abundance is artificially high. 

\begin{figure}
   \begin{center}
     \centering
     \includegraphics[width=\textwidth]{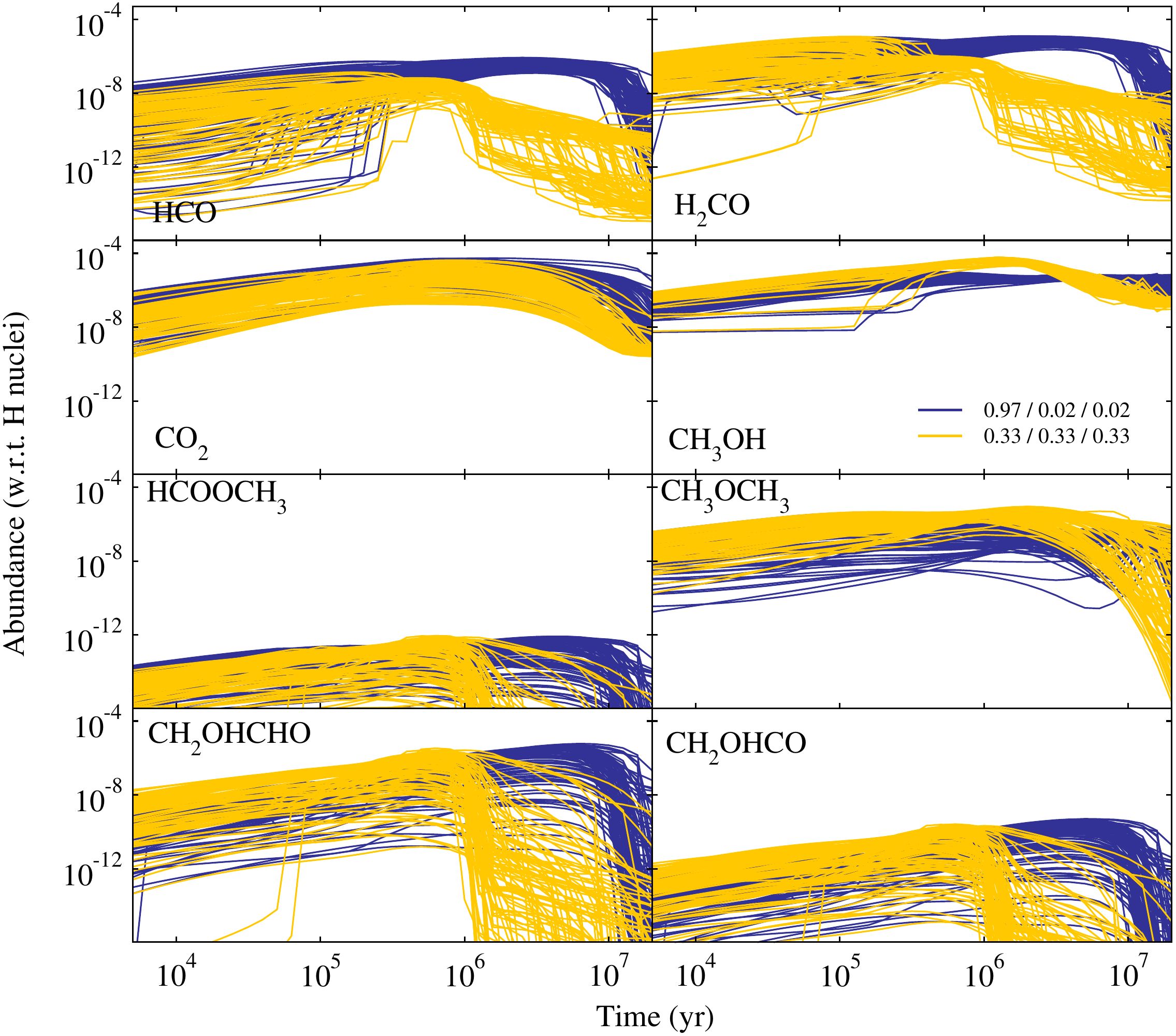}
      \caption{Evolution of ice abundances according to two different branching ratios: 0.97~/~0.02~/~0.02 
      (standard network) and 0.33 / 0.33 / 0.33 (updated network). 
      Results using the standard network are shown in blue and using the updated network in yellow.
      Abundances much lower than 10$^{-16}$ are negligible and are therefore not shown here.  Figure reproduced from \citet{Penteado:2017}.}
        \label{diff_rates}
   \end{center}
   \end{figure}

Using the standard network, its main formation reaction \ce{H + H2CO -> HCO + H2} is much more efficient in the model (97~\%) than \ce{H + H2CO -> CH3O} or \ce{H + H2CO -> CH2OH} which equally share the remainder. This is because the barriers for the different channels originate from different sources (gas phase data vs.~surface experiments). Surface experiments do not support these extreme branching ratios on the surface. In a recent experimental study \citet{Chuang:2016} investigated the hydrogenation of \ce{H2CO}. They  indeed observe the addition reaction to \ce{CH3O} and the abstraction to \ce{H2 + HCO}, and they cannot exclude the channel to form \ce{CH2OH}. Here, we alter the network by adopting equal ratios for all 
three channels and perform 100 additional simulations using this network. The evolution of the 
ice abundance of the species involved in the hydrogenation of CO is shown in Figure~\ref{diff_rates} in yellow. A comparison 
is made between the original 100 runs of our standard network in blue.

The timescale of conversion of HCO to \ce{H2CO} and \ce{CH3OH} is reduced by a few million years, 
since there are less back reactions. Moreover, as a consequence of the reduction in HCO, an increase in the methanol abundance can be observed, 
since \ce{CH3O} and \ce{CH2OH} will now predominantly react with H to form methanol. For the more complex species, both a reduction in the peak intensity and a change in the formation timescale can be observed. 

In summary, the sensitivity analysis allows us to expose connections between different species in the total reaction network. It shows the importance of the \ce{H2CO + H} reaction. By changing the branching ratios of this single reaction to more realistic values, the timescale and peak abundances of many COMs have changed. Several other reactions with branching ratios exist in the network and we advise a complete scrutinization of the whole network. 
The overproduction of CH$_4$ by the model with respect to observations could be due to too efficient destruction reactions of COMs, because of inaccuracy in branching ratios of reactions leading to CH$_4$ or its precursors.

\section{Role of exothermicity in non-thermal desorption and diffusion}
As a last example, we would like to present Molecular Dynamics simulations to examine the role of exothermicity in non-thermal desorption and diffusion. In our earlier example of low temperature COM formation, we already showed the importance of exothermicity in the build-up of ice mantles. Here we would like to quantify the outcome of exothermicity. Excess energy can be in the form of translational, rotational, or vibrational energy or electronic excitation. How this energy is spread is highly system dependent\citep{Polanyi:1986}; it does not only depend on the specific reaction but also on the reaction environment and the configuration in which it occurs. The excitation energy could be applied for desorption, but also for restructuring of the grain mantle as shown above, or diffusion of the species.  Reactive or chemical desorption \citep{Garrod:2006b,Garrod:2007,Dulieu:2013} is hence only one possible outcome. In analogy, we introduce the concept of chemical or non-thermal diffusion where the energy released during the surface reaction allows the product to diffuse over the surface with the possibility to meet another reactant for subsequent reactions \citep{Arasa:2010,Lamberts:2014I}. In this way, the excess energy of one reaction would enable subsequent reactions. We expect the type of excitation to have an effect on the outcome. 

Molecular Dynamics (MD) simulations follow Newton's equations of motion and allow one to keep track of the energy, potential or kinetic energy, by giving a species some initial energy. KMC models, although microscopic, do not offer this opportunity. Here we aim to investigate the fate of kinetically excited species on the surface of an ice mantle. We consider an additional kinetic energy between 0.5 and 5 eV to cover the full range from reactions with a small exothermicity to photodissociation reactions where the excitation energy after bond breaking can be several eV. Crystalline water ice is used as a substrate and three different surface species are considered, \ce{CO2}, \ce{H2O} and \ce{CH4}, which are among the most common ice species \citep{Boogert:2015}. Moreover, these species span a range in binding energies, number of internal degrees of freedom and molecular weight. In a later stage, we would also like to study the effect of other types of excitation, especially vibrational and rotational as well as the effect of the type of substrate (amorphous versus crystalline).

We roughly apply the following procedure: First the different binding sites on the substrate are identified by deposition of an admolecule at different positions on the water surface. Next, the strongest binding sites are selected, and from each site new simulations are issued where the admolecule is given a specific amount of additional translational kinetic energy in a random direction. Finally, the outcome of the simulation is detected by the software and stored for post-analysis. We classify each simulation into: desorption, adsorption, penetration, FAIL, and TIME LIMIT. For more details, we refer to \citet{Fredon:sub}.

\begin{figure*}[t!]
  \includegraphics[width=\textwidth]{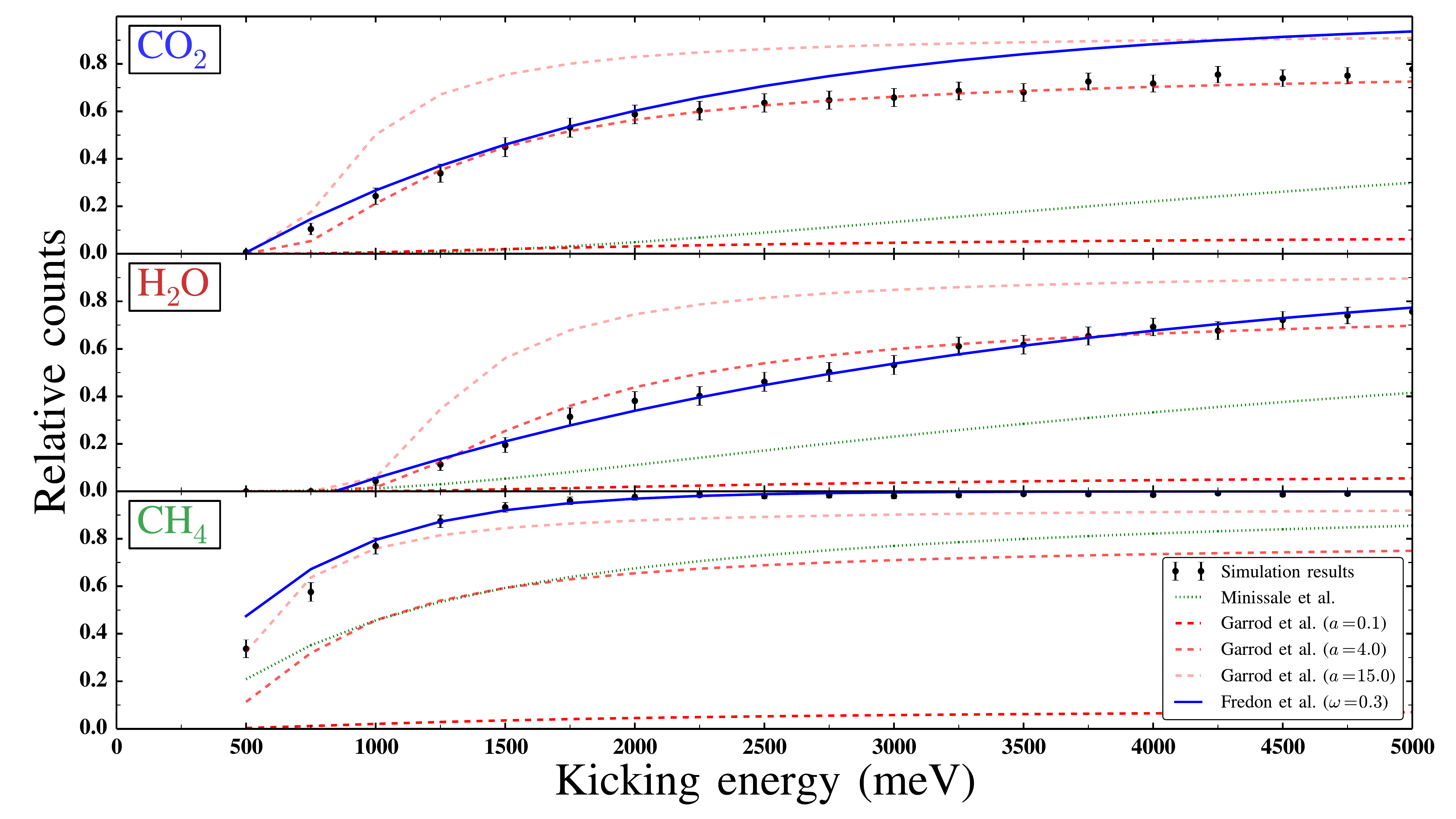}
  \caption{Desorption probability as a function of the initial translational energy given to the admolecule. The dots represent the data from our simulations with an error bar of $3 \sigma$. The blue lines, representing our model, show the closest agreement with the simulation data.  Figure reproduced from \citet{Fredon:sub}.}
  \label{fig: Outcomes}
\end{figure*}

For all three admolecules, desorption or adsorption are the most likely outcomes; penetration is found to occur in only a few simulations, all for high kicking energies. 
Figure~\ref{fig: Outcomes} shows the fraction of simulations resulting in desorption as a function of the initial translational energy given to the three admolecules. Each fraction is based on 1,400 individual simulations. As expected, the desorption probability increases with increasing kick energies. There appears to be a threshold energy required for desorption, which coincides with the binding energy of the different species to the surface. The dashed and dotted curves represent models for chemical desorption that are generally used. Both the model by \citet{Garrod:2007} and the model \citet{Minissale:2016} cannot reproduce our simulation results with one universal expression. The model by  \citet{Garrod:2007} focuses on vibrational excitation instead of translational excitation which explains the discrepancy. The model by \citet{Minissale:2016}, on the other hand, does not reproduce the threshold value. We find our data to be best reproduced by 
\begin{equation}
  P_\text{CD}(E_\text{react}^\text{trans}) = 1 - \exp\left(-\omega \frac{E_\text{react}^\text{trans} - |E_\text{bind}|}{|E_\text{bind}|}\right) 
  \label{equ: CD Adrien}
\end{equation}
with $\omega =0.3$. This is shown by the blue solid line in Figure~\ref{fig: Outcomes}.

As an example we will consider reactive desorption in two specific reactions. We start with 
\begin{equation}
 \ce{H + OH -> H2O}.
\end{equation}
This reaction has an enthalpy of reaction of 4.89~eV. If we assume an equipartition of the energy over all 9 degrees of freedom ($E_\text{react}^\text{trans} = 3 \times 1/9 \times 4.89 = 1.63$ eV), the desorption probability is
\begin{equation}
  P_\text{CD} = 1 - \exp\left(-0.3 \frac{1.630 - 0.839}{0.839}\right) = 0.24.
\end{equation}
Experiments on non-porous ASW give a desorption efficiency of 0.30 $\pm$ 0.10 \citep{Minissale:2016} and our values are hence in quite good agreement. 
We can make a similar calculation for \ce{CH3 + H -> CH4} which has a heat of formation of 4.55~eV. The chemical desorption probability from a crystalline surface is now 0.75. This significantly higher efficiency of \ce{CH4} desorption, versus the other molecules, is very interesting. Many chemical models of ice mantle production appear to overproduce solid-phase methane, as compared to astronomical measurements (as a fraction with respect to water)\citep{Garrod:2006,Oberg:2008}, including our simulations presented in Section 3.

\begin{figure*}[t!]
  \begin{minipage}[b]{0.5\textwidth}
    \centering
    \includegraphics[width=1.1\textwidth]{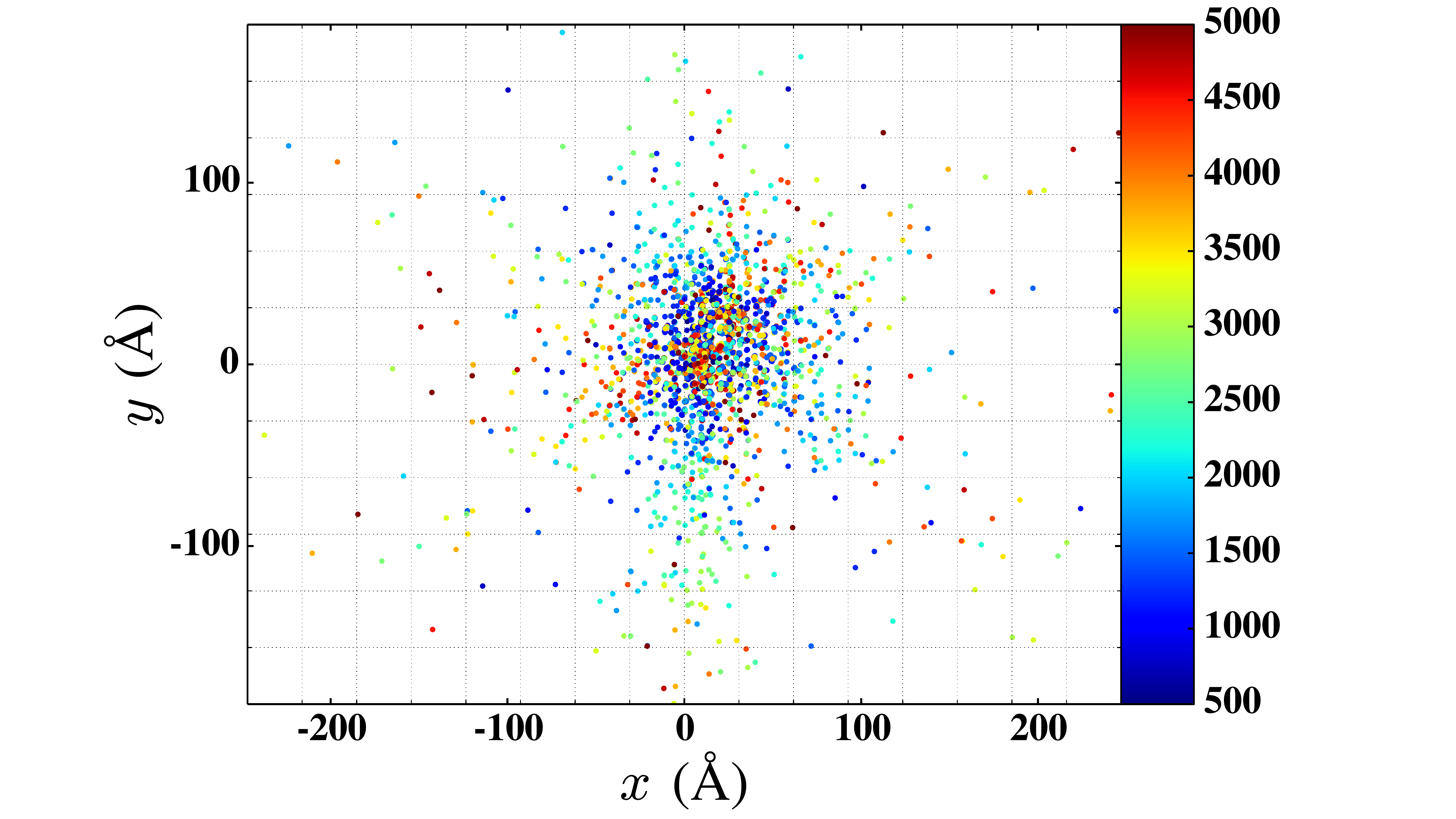}
    (a)
  \end{minipage}
  \hfill
  \begin{minipage}[b]{0.5\textwidth}
    \centering
    \includegraphics[width=1.1\textwidth]{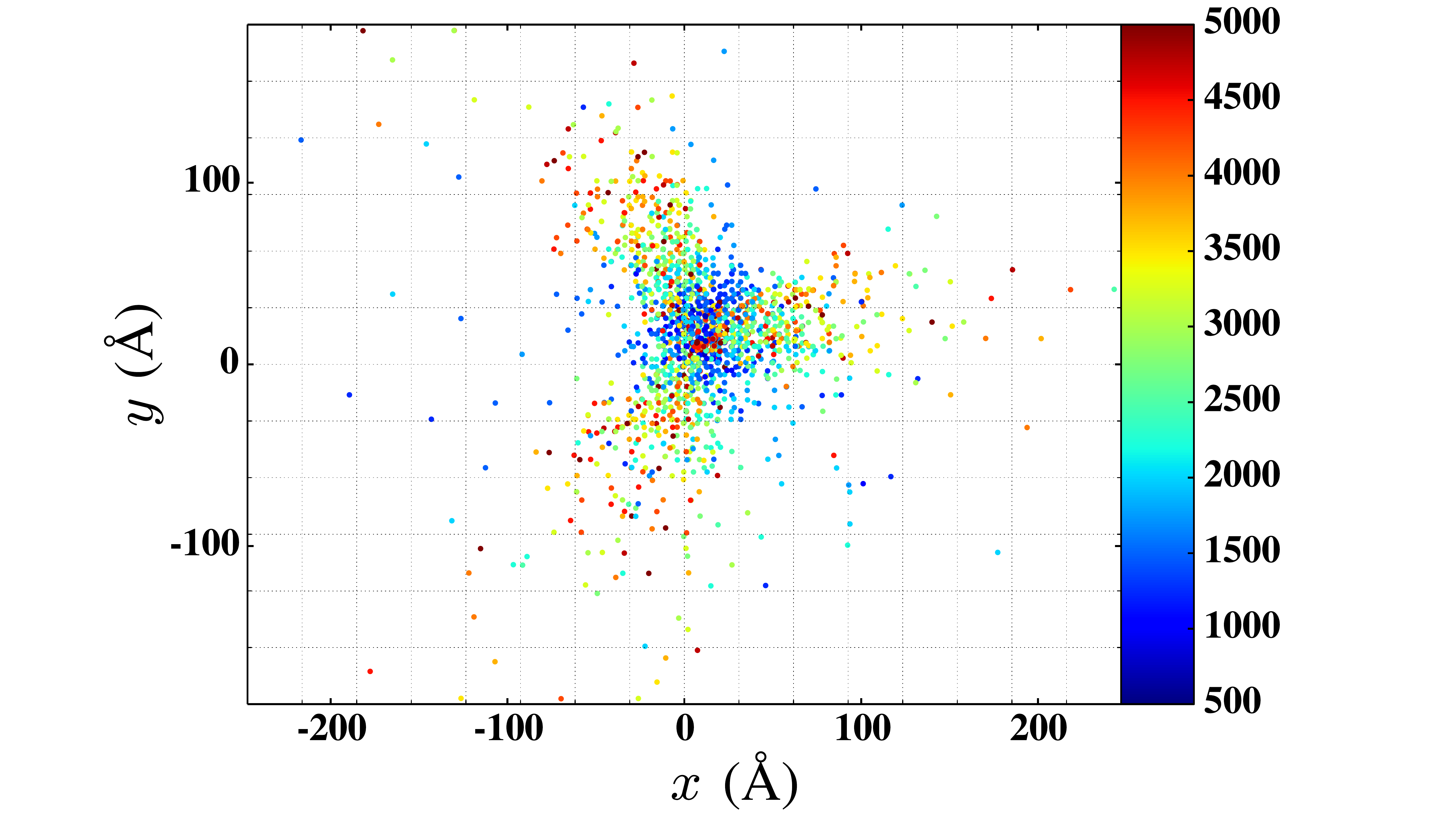}
    (b)
  \end{minipage}
  \\
  \begin{minipage}[c]{1.\textwidth}
    \centering
    \includegraphics[width=0.55\textwidth]{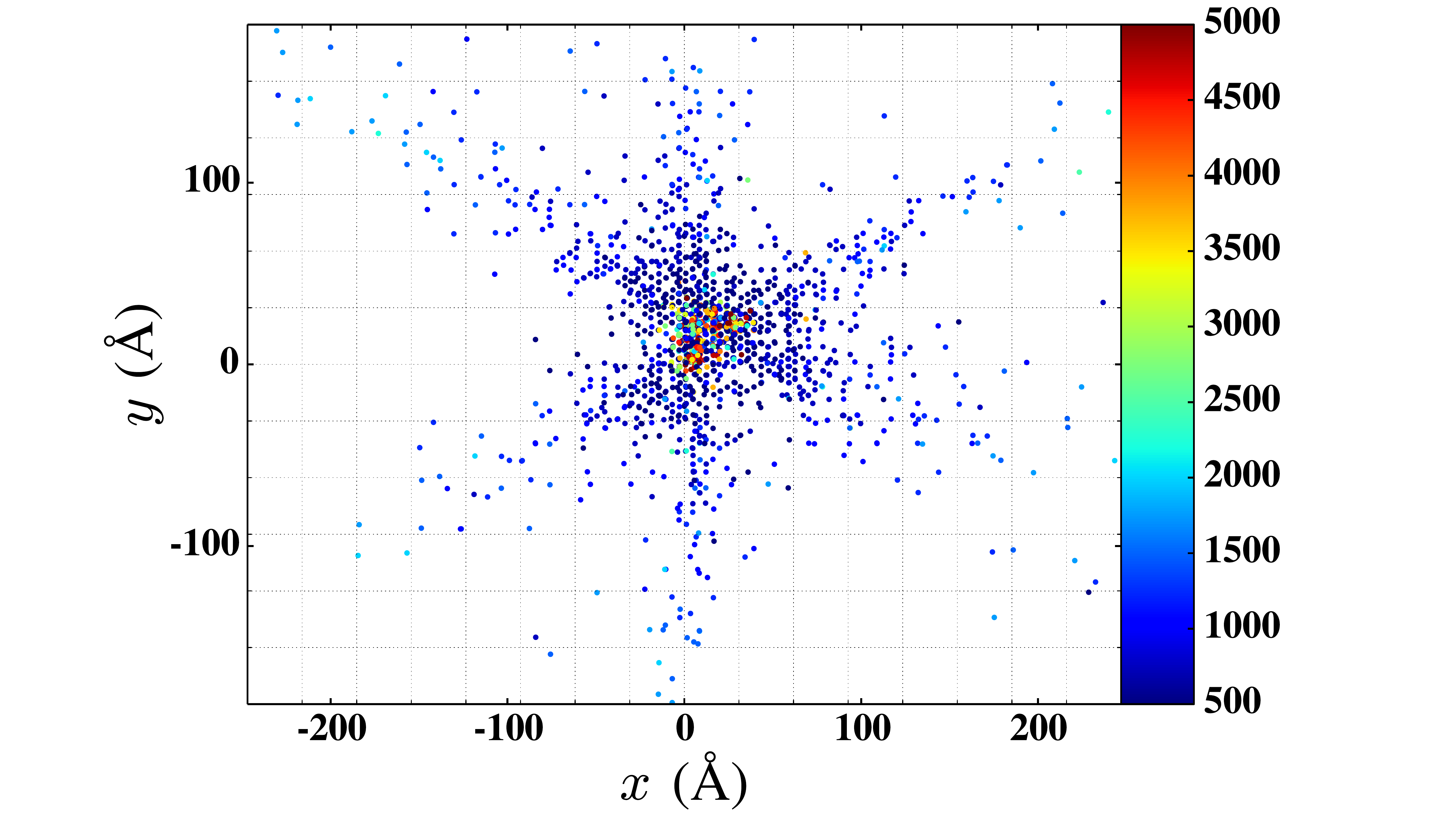}
    \vfill
    (c)
  \end{minipage}
  \caption{Top view of the simulation cells (original and periodic copies) indicating the traveled distance and direction of the admolecules for each ``adsorption'' trajectory. The original simulation cell is the one in the center and the copies are represented by the grid. The colored dots represent the positions of the \ce{CO2} (a), \ce{H2O} (b), and \ce{CH4} (c) admolecules at the end of the energy dissipation simulations. The colorbar represents the initial translational energy of the admolecules. Figure reproduced from \citet{Fredon:sub}.}
  \label{fig: Traveled distance unwrapped kick energy}
\end{figure*}

Figure~\ref{fig: Traveled distance unwrapped kick energy} indicates the traveled distance and direction of the admolecules for each ``adsorption'' trajectory. The figures shows multiple copies of the same simulation cell stacked in both $x$ and $y$ direction; the individual cells are indicated by the grid. All admolecules have started from the central simulation box and they traveled through the periodic boundary conditions (in some cases multiple times) to their final position indicated by the dots. The dots are color coded according to their initial kicking energy. The blue dots dominate the \ce{CH4} panel because trajectories with higher initial translational energies will have resulted in desorption of the species and are hence not included in the plot. For many trajectories, the admolecule has traveled a substantial distance before coming to a stand still, up to 1157~{\AA} for \ce{H2O} (not shown in the figure). 

The \ce{CH4} results in panel (c) in Fig.~\ref{fig: Traveled distance unwrapped kick energy} show a strong angular dependence following the hexagonal pattern of the underlying crystalline substrate. Collisions of the \ce{CH4} molecule with dangling protons lead either to redirection of the molecule through the ``hexagonal channels'' on the surface or to desorption of the molecule. In contrast, \ce{H2O} prefers to interact with the dangling protons of the ice and three main directions in which \ce{H2O} travels on the surface are observed, following the dangling proton pattern. Indeed, as can be seen in Fig.~\ref{fig: Traveled distance unwrapped kick energy}, these directions do not coincide with the hexagonal channels of \ce{CH4}. \ce{CO2} appears to be less affected by surface protrusions. 

\section{Conclusions}
In conclusion, we have shown how different computational chemistry techniques can be applied to obtain more information about grain surface chemistry. We have used three different examples of different problems that required three very different computational techniques. To scrutinize a large gas-grain chemistry network through a sensitivity analysis requires a simulation method that can handle large reaction networks at very low computational costs to allow for 10,000 simulations. Currently, this is limited to the use of macroscopic methods, which we know not to capture all surface chemistry details. In this case, it exposed a weakness in the network: the uncertainty in the branching ratio of \ce{H + H2CO}. We hope that this will be followed up experimental groups.

The low temperature formation route of glycoaldehyde does not require diffusion or external UV. HCO radicals should simply form in close proximity. This required a microscopic, statistical approach. With our KMC model, we were indeed able to show that these complex molecules can form a temperatures as low as 10~K. It further showed the importance of exothermicity in the building of the ice layer. Again a microscopic method was required to obtain this type of information. 

Finally, using Molecular Dynamics simulations we could quantify the result of exothermicity in the form of kinetic excitation. Species can indeed travel a substantial distance before thermalizing or kinetic excitation can result in non-thermal desorption. However, the picture may change when also including vibrational and rotational excitation and using an amorphous substrate.

\end{document}